\documentclass[aps,prl,twocolumn,groupedaddress,amsmath,superscriptaddress]{revtex4-1}

\usepackage[T1]{fontenc}
\usepackage[latin9]{inputenc}
\usepackage{dcolumn}% Align table columns on decimal point
\usepackage{amsmath}
\usepackage{amssymb}
\usepackage{amsfonts}
\usepackage{graphicx}
\usepackage{hyperref}
\usepackage{color}
\usepackage{mathrsfs}
\usepackage{multirow}
\usepackage{float}
\usepackage{booktabs}
\usepackage{bm}
\usepackage{dsfont}
\usepackage{txfonts}
\usepackage[english]{babel}
\usepackage{supertabular}
\definecolor{nblue}{rgb}{0.3,0.3,1.0}%229
\definecolor{ngreen}{rgb}{0.2,0.7,0.2}%161
\definecolor{nred}{rgb}{0.9,0.1,0}%711&900
\definecolor{nblack}{rgb}{0,0,0}

\usepackage[mathlines]{lineno}% Enable numbering of text and display math
\modulolinenumbers[5]% Line numbers with a gap of 5 lines
\begin{document}

\title{Versatile Multipartite Einstein-Podolsky-Rosen Steering via a Quantum Frequency Comb}

\author{Yin~Cai}
\email{caiyin@xjtu.edu.cn}
\address{Key Laboratory for Physical Electronics and Devices of the Ministry of Education $\&$ Shaanxi Key Lab of Information Photonic Technique, Xi'an Jiaotong University, Xi'an 710049, China}
\address{Laboratoire Kastler Brossel, Sorbonne Universit\'e, ENS-PSL Research University, Coll\`ege de France, CNRS; 4 place Jussieu, F-75252 Paris, France}
\author{Yu~Xiang}
\email{xiangy.phy@pku.edu.cn}
\address{State Key Laboratory for Mesoscopic Physics, School of Physics, Frontiers Science Center for Nano-optoelectronics, $\&$ Collaborative Innovation Center of Quantum Matter, Peking University, Beijing 100871, China}
\address{Beijing Academy of Quantum Information Sciences, Haidian District, Beijing 100193, China}
\address{Collaborative Innovation Center of Extreme Optics, Shanxi University, Taiyuan, Shanxi 030006, China}
\author{Yang~Liu}
\address{Key Laboratory for Physical Electronics and Devices of the Ministry of Education $\&$ Shaanxi Key Lab of Information Photonic Technique, Xi'an Jiaotong University, Xi'an 710049, China}
\address{State Key Laboratory of Transient Optics and Photonics, Xi'an Institute of Optics and Precision Mechanics, Chinese Academy of Sciences, Xi'an 710119, China}
\address{University of Chinese Academy of Sciences (UCAS), Beijing 100049, China}
\author{Qiongyi~He}
\email{qiongyihe@pku.edu.cn}
\address{State Key Laboratory for Mesoscopic Physics, School of Physics, Frontiers Science Center for Nano-optoelectronics, $\&$ Collaborative Innovation Center of Quantum Matter, Peking University, Beijing 100871, China}
\address{Beijing Academy of Quantum Information Sciences, Haidian District, Beijing 100193, China}
\address{Collaborative Innovation Center of Extreme Optics, Shanxi University, Taiyuan, Shanxi 030006, China}
\author{Nicolas~Treps}
\address{Laboratoire Kastler Brossel, Sorbonne Universit\'e, ENS-PSL Research University, Coll\`ege de France, CNRS; 4 place Jussieu, F-75252 Paris, France}

\begin{abstract}
Multipartite Einstein-Podolsky-Rosen steering is an essential resource for quantum communication networks where the reliability of equipment at all of the nodes cannot be fully trusted. Here, we present experimental generation of a highly versatile and flexible repository of multipartite steering using an optical frequency comb and ultrafast pulse shaping. Simply modulating the optical spectral resolution of the detection system using the pulse shaper, this scheme is able to produce on-demand  4, 8 and 16-mode Gaussian steering without changing the photonics architecture. We find that the steerability increases with higher spectral resolution. For 16-mode state, we identify as many as 65 534 possible bipartition steering existing in this intrinsic multimode quantum resource, and demonstrate that the prepared state steerability is robust to mode losses. Moreover, we verify four types of monogamy relations of Gaussian steering and demonstrate strong violation for one of them. Our method offers a powerful foundation for constructing quantum networks in real-world scenario.\\

\end{abstract}
\maketitle

% different types of entanglement are required for different applications...a special type of entanglement called quantum steering has attracted much attention as a resource for one-sided device-independent quantum cryptography.
In 1935, Einstein, Podolsky, and Rosen (EPR)~\cite{EPR35} pointed out one of the
most counter-intuitive features of quantum mechanics: the local measurements performed on one system $A$ seem to immediately affect the state of another distant system $B$. The nonlocal effect embodied in this EPR paradox was named `steering' by Schr\"odinger~\cite{Schrodinger35}. In 2007, the concept of steering was rigorously defined~\cite{Howard07PRL,Howard07PRA} in view of local hidden state model, and it was specified that steering allows for the verification of entanglement shared between two remote parties even if one of them is untrusted~\cite{ReidRMP, Eric13,cavalcanti17review}. Different to Bell nonlocality~\cite{BrunnerRMP} and entanglement~\cite{entRMP} where the roles of involved parties are symmetric, steering is a directional form of nonlocality for which the losses or noises can act asymmetrically~\cite{one-way-Theory, He15,ReidJOSAB}. For two party communication, steering has been applied to realize secure quantum teleportation~\cite{SQT13Reid,SQT15,SQT16_LiCM}, one-sided device-independent quantum cryptography~\cite{1sDIQKD,1sDIQKD_howard,HowardOptica,CV-QKDexp} and subchannel discrimination~\cite{subchannel,subchannel16}. 

When one aims at realistic applications in quantum communication network where the reliability of devices or the possibility of dishonest measurement are critical, multimode EPR steering is relevant~\cite{genuine13,Adesso15,ANUexp}. For instance, if for a certain quantum task a given form of multipartite entanglement is required, multipartite steering allows one to verify that this entanglement is present without the need of the full trust of all measurement devices~\cite{GiannisQSS}. Recently, experimental generation of multipartite steering in the continuous variable regime has developed rapidly. Armstrong \emph{et.al.}~\cite{ANUexp} and Deng \emph{et.al.} ~\cite{prlSu} produced multipartite steering by mixing squeezed lights via beam splitter networks. These experiments, however, require re-modifying optical setups for creating different steerable states, and in particulars need more squeezed light sources and beamsplitters to improve the steerability and enlarge the network, thus lacking versatility and flexibility. In contrast, via parametric down conversion of a train of ultrafast pulses, multiple frequencies can be quantum correlated simultaneously to generate multimode entanglement from a single cavity~\cite{Pinel,Pfister}, which allows on-demand generation of multipartite entangled states without changing the optical circuit. This technique offers a convenient and powerful platform for realizing many squeezed spectro/temporal modes copropagating within a single beam~\cite{Roslund}. It produces full multipartite entanglement of up to 115 974 possible nontrivial mode partitions for a 10-mode system~\cite{multient2015, vogel2016}, as well as creates on-demand cluster networks~\cite{Cainc}. 
%These pioneer works inspire us to explore whether this multimode quantum platform can generate more stringent forms of correlations such as EPR steering, then investigate the properties of prepared multipartite steerable state.

In this letter, we demonstrate multipartite steering between partitions of such an intrinsically multimode resource generated from the parametric down conversion of an optical frequency comb. The many different steering configurations are obtained without changing the optical setup but simply projecting the state on the required spectral modes, either with a pulse shaped homodyne detection or post-processing the data from the fully spectrally resolved measurement. We find that the steerability increases when one increases the spectral resolution of the measurement, decomposing the signal spectrum into four-, eight-, and sixteen- mode states. We also show up to 65 534 steering bipartitions co-existing in the multimode quantum system and demonstrate their robustness to the loss of modes. Moreover, we examine the monogamy relations of Gaussian steering, which qualitatively defines the information security among different parties,  and show the possible violation of monogamy inequalities. Our results indicate that this multimode quantum platform is a scalable and versatile resource for multipartite steering.

%\section{Reconstructing multimode gaussian states}
%\blu{1. Set-up: comb and shaping, squeezed eigenmodes}
%\blu{2. covariance matrix definition}
%\blu{3. criteria}
%
\begin{figure}[tbp]
	\includegraphics[width=1\linewidth]{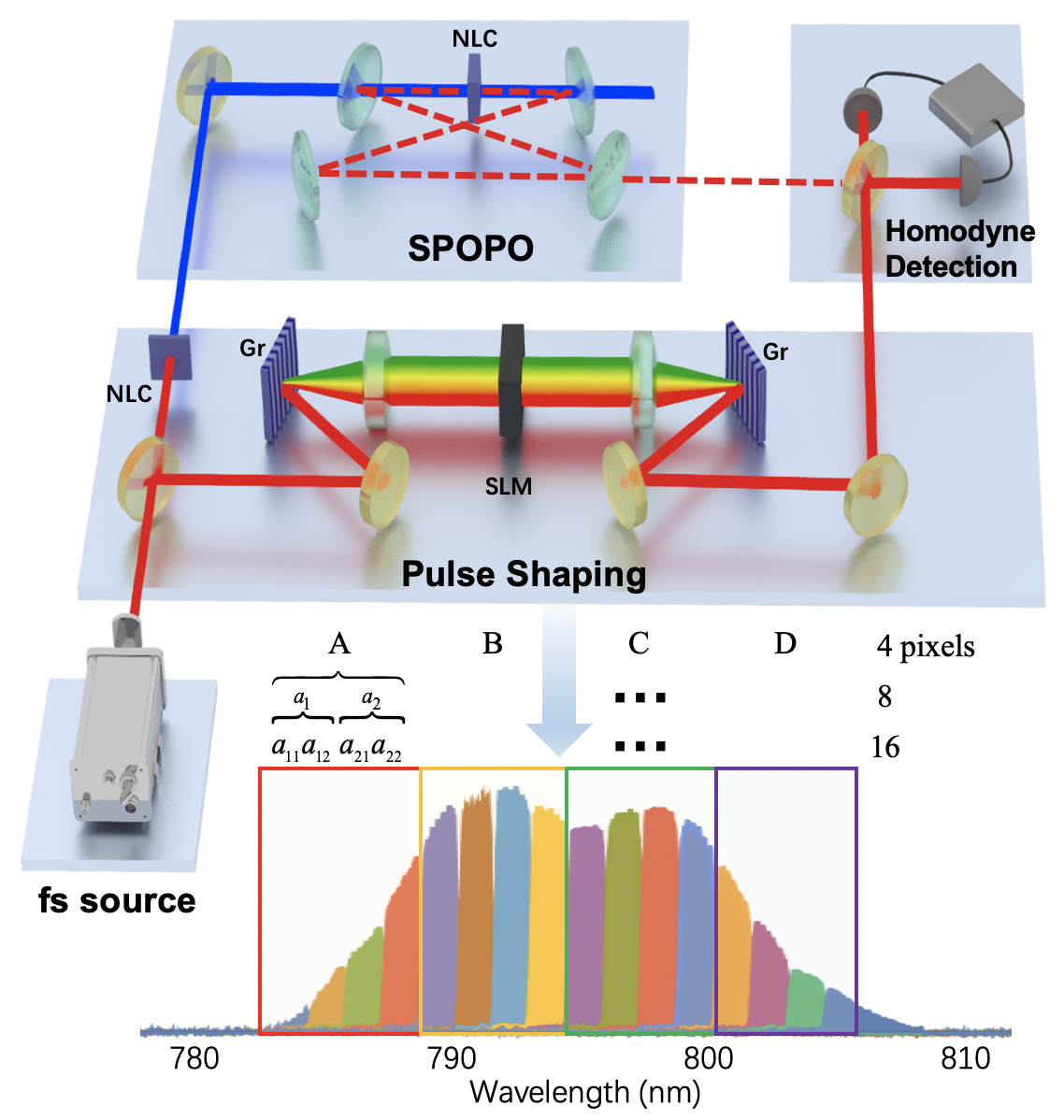}
	\caption{\label{fig1} Experimental set-up of the multimode quantum resource via synchronously pumping an optical parametric oscillator (SPOPO). A Homodyne detection with pulse shaping on the local oscillator (LO) is applied. The bottom shows the pulse shaping on the LO with controllable spectral resolution. The spectrum of the LO is divided into 4, 8 and 16 spectral bands, respectively. NLC represents a nonlinear crystal, Gr is a grating, and SLM means a spatial light modulator.}
	\label{fig1}
\end{figure}
\textit{Experimental realization.} 
As illustrated in Fig.~\ref{fig1}, a femtosecond optical frequency comb, composed of around $10^5$ frequencies centered at 795 nm and with a 76MHz spacing, is frequency doubled and pumps an optical parametric oscillator (SPOPO) below threshold. Within the non-linear crystal, pairs of photons are generated whose frequency sum matches one of the pump frequency, due to energy conservation. Furthermore, synchronous cavity ensures that the output field is also a coherent optical frequency comb, matching the source laser spectral properties. The result of this downconversion process thus forms a multimode quantum frequency comb, with a highly rich correlation pattern in the frequency domain, demonstrated to be fully multipartite~\cite{multient2015}. This same state can also be described as a coherent superposition of many co-propagating, independently squeezed spectro/temporal modes. This quantum state is interrogated with an homodyne detection, in which ultrafast pulse shaping  is employed to select the spectral amplitude of the local oscillator (LO), and thus the measured spectro/temporal mode~\cite{Roslund}.

The bottom plot in Fig.\ref{fig1} shows how the spectrum of the LO is partitioned: sixteen spectral bands of equal frequency width are used. Scanning the relative phase between the quantum frequency comb and the LO, the amplitude $\hat{x}$ and phase $\hat{p}$ quadrature of the mode selected by the LO are measured. Shaping the LO into pairs of spectral bands, the correlations are also measured. Note that in our case the cross correlations of the forms $\langle\hat{x}\hat{p}\rangle$ or $\langle\hat{p}\hat{x}\rangle$ are measured to be negligible. From these measurements, the full covariance matrix can be revealed, as was demonstrated in \cite{Roslund}. In the present case, one can show that the eigenmodes of this covariance matrix display squeezing up to $\sim$-5dB (see details in Appendix~\cite{supp}).  

\textit{Covariance matrix (CM) and steerability.} As explained above, the covariance matrix is originally measured in a basis of  16 equally spaced spectral band modes. Defining $\hat{\vec \zeta} = (\hat x_1, \hat p_1, \ldots, \hat x_i, \hat p_i, \ldots)$ where $\hat x_i$ and $\hat p_i$ are the quadrature operators of the i$^\textrm{th}$ spectral band, the covariance matrix elements are formally defined as $\frac12\left\langle {{{\hat\zeta }_i}{{\hat \zeta }_j} + {{\hat \zeta }_j}{{\hat \zeta }_i}} \right\rangle - \left\langle {{{\hat \zeta }_i}} \right\rangle \left\langle {{{\hat \zeta }_j}} \right\rangle$. From this matrix, it is possible to derive the covariance matrix of any bipartition, where each party can be composed of several modes. Let us consider that the first party is composed of $m$ modes and the second of $n$ modes, the covariance matrix can be recasted in the form $\sigma_{mn} = \left( \begin{array}{ccc} \mathcal{M} & \mathcal{C}  \\ \mathcal{C}^\top & \mathcal{N}  \end{array} \right)$ where  submatrices $\mathcal{M}$, $\mathcal{N}$ correspond to the reduced states of each party respectively, and submatrix $\mathcal{C}$ represents the correlation between them. 

In practice, we reconstruct the CMs with four-, eight- and sixteen modes, as plotted at the bottom of Fig.~\ref{fig1}. For instance, the spectral component labeled with A is further divided to $\{a_1, a_2\}$, and then $\{a_{11}, a_{12}, a_{21}, a_{22}\}$, and same operation for the spectral bands B, C and D~\cite{supp}. The CMs of four-, eight- and sixteen-mode states correspond to projecting the same state of the quantum frequency comb onto the spectral band basis with increasing measurement resolution in the pulse shaping process, without changing any setting of the original resource. The steerability from party $m$ to party $n$ is quantified by~\cite{Adesso15}
\begin{equation}
\mathcal{G}^{m \to n}(\sigma_{mn})=\mathrm{max}\left\{ {0, - \sum\limits_{i:\bar \nu_i^{mn/m} < 1} {\ln (\bar \nu_i^{mn/m})} } \right\},
\label{eq2}
\end{equation}
where $\{\bar \nu_i^{mn/m}\}$ are the symplectic eigenvalues of  $\bar\sigma_{mn/m}=\mathcal{N}-\mathcal{C}^\top\mathcal{M}^{-1}\mathcal{C}$, which is the Schur complement of partition $m$. Under Gaussian local operations and classical communication, this quantity $\mathcal{G}^{m \to n}$ is always monotone.  A zero $\mathcal{G}^{m \to n}$ means that $m$ is not able to steer $n$ by Gaussian measurements, but it may succeed with non-Gaussian measurements~\cite{Yunon}.

%\section{Results}

%\blu{1. EPR steering VS resolution:(entanglement with fixed spectral components %always)}\\
%\blu{(1) 4$<$8$<$16 steering is better with increased resolution of measurement. seen in table\\
%(2) why? more squeezed modes are extracted with higher resolution. \\
%(3) Other OPO system doesn't have this property. Comparison (Supplementary) \\
%(4) Explain the jump of points.\\
%(5) Asymmetrically controllable.}
\begin{figure}[b]
	\centering
	\includegraphics[width=0.95\linewidth]{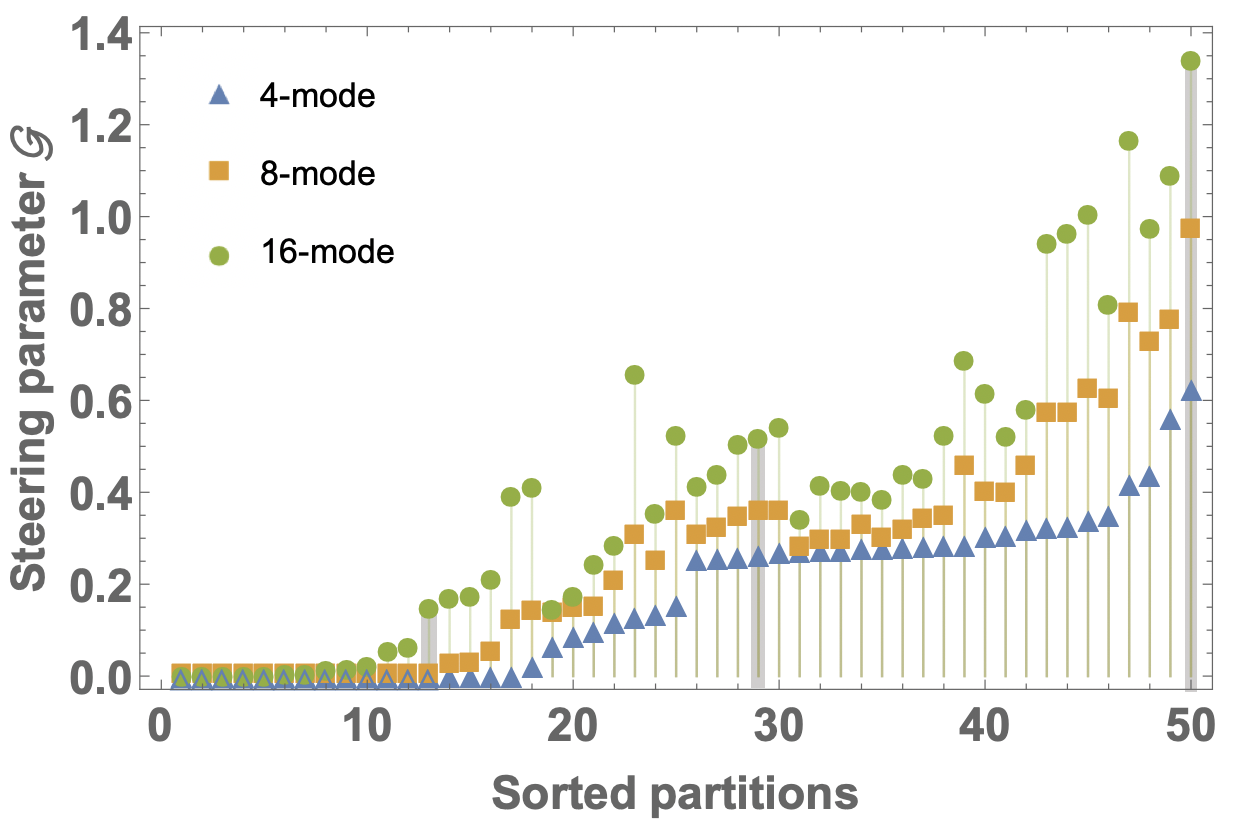}
	\caption{Multipartite EPR steering among four spectral bands $ABCD$ are measured experimentally with 4 (blue triangles), 8 (orange squares), 16 (green dots) spectral pixels, respectively. The partitions are arranged from small to large according to the steerability of 4-mode state. Marks on the same vertical line correspond to the steering between same spectral bands with different spectral pixels.} \label{fig2}
\end{figure}

\textit{Results.} Figure~\ref{fig2} shows multipartite steering among the four spectral bands $ABCD$ when the whole spectrum is divided into 4 (blue triangles), 8 (orange squares) or 16 (green dots) pixels. Three marks on the same vertical line correspond to the bipartition steerability between identical spectral bands but measured with different resolutions. Note that here we show all possible nontrivial partitions for the four-mode state, while for comparison we only show the partitions of eight- and sixteen-mode states that are correspondingly subdivided from the four spectral bands $ABCD$ without taking into account the mixtures among them. Three representative groups (gray slices) are delineated in Table \ref{table1}. Interestingly, although the spectral components are fixed, the steerability raises sharply with the increase of measurement resolutions for all possible partitions. For instance,  $\mathcal{G}^{(AB) \to C}<\mathcal{G}^{(a_1,a_2,b_1,b_2) \to (c_1,c_2)}<\mathcal{G}^{(a_{11},a_{12},a_{21},a_{22},b_{11},b_{12}, b_{21},b_{22})\to (c_{11},c_{12},c_{21},c_{22}})$. %especially when the steering party or the steered party comprises more than one mode. (% we try to explain the jump points when A->BCD, AB->CD).   
This phenomenon can have several origins. First, increasing the resolution, one better resolves the spectral shape of the eigenmodes, hence measured squeezing increases and thus the steerability. Furthermore, adding more parties increase the number of extracted modes, parameter which also influences  the steerability as we will study in the last section of this paper on monogamy. Hence, we see that our scheme allows for the preparation of versatile multipartite EPR steering and simultaneously increase steerability simply modulating the spectral resolution of the detection system, i.e. acting on the spectral shape of the LO. Moreover, we can generate richer steering structures by asymmetrically adjusting the pulse shaping. For example, increasing the resolutions either on the steering party or steered party, one-way steering is realized~\cite{supp}.  
\begin{table}[b]
	\caption{\label{table1}
		Steerability with different partitions (one-to-multi, multi-to-one and multi-to-multi) and their respective spectral resolutions.}
	\begin{ruledtabular}
		\renewcommand\arraystretch{1.2}
		\begin{tabular}{ccc}
			\textrm{Resolution}&
			Partitions&
			{$\mathcal{G}$}\\
			\colrule
			4 & $\mathcal{G}^{A \to (BC)}$ & 0 \\
			8 & $\mathcal{G}^{(a_1, a_2) \to (b_1, b_2, c_1, c_2)}$ & 0 \\
			16 & $\mathcal{G}^{(a_{11}, a_{12}, a_{21}, a_{22}) \to (b_{11}, b_{12}, \ldots, c_{21}, c_{22})}$ & 0.1481$\pm$0.0714 \\
			\hline
			4 & $\mathcal{G}^{(AB) \to C}$ & 0.2633$\pm$0.0881 \\
			8 & $\mathcal{G}^{(a_1, a_2, b_1, b_2) \to (c_1, c_2)}$ & 0.3537$\pm$0.0927
			\\
			16 & $\mathcal{G}^{(a_{11}, a_{12}, \ldots, b_{21}, b_{22}) \to (c_{11}, c_{12}, c_{21}, c_{22})}$ & 0.5177$\pm$0.1139
			\\
			\hline
			4 & $\mathcal{G}^{(CD) \to (AB)}$ & 0.6226$\pm$0.0963\\
			8 & $\mathcal{G}^{(c_1, c_2, d_1, d_2) \to  (a_1, a_2, b_1, b_2)}$ & 0.9692$\pm$0.1372
			\\
			16 & $\mathcal{G}^{(c_{11}, c_{12}, \ldots, d_{21}, d_{22}) \to (a_{11}, a_{12}, \ldots, b_{21}, b_{22})}$ & 1.3415$\pm$0.1836
			\\
		\end{tabular}
	\end{ruledtabular}
\end{table}

\begin{figure}[tb]
	\includegraphics[width=0.85\linewidth]{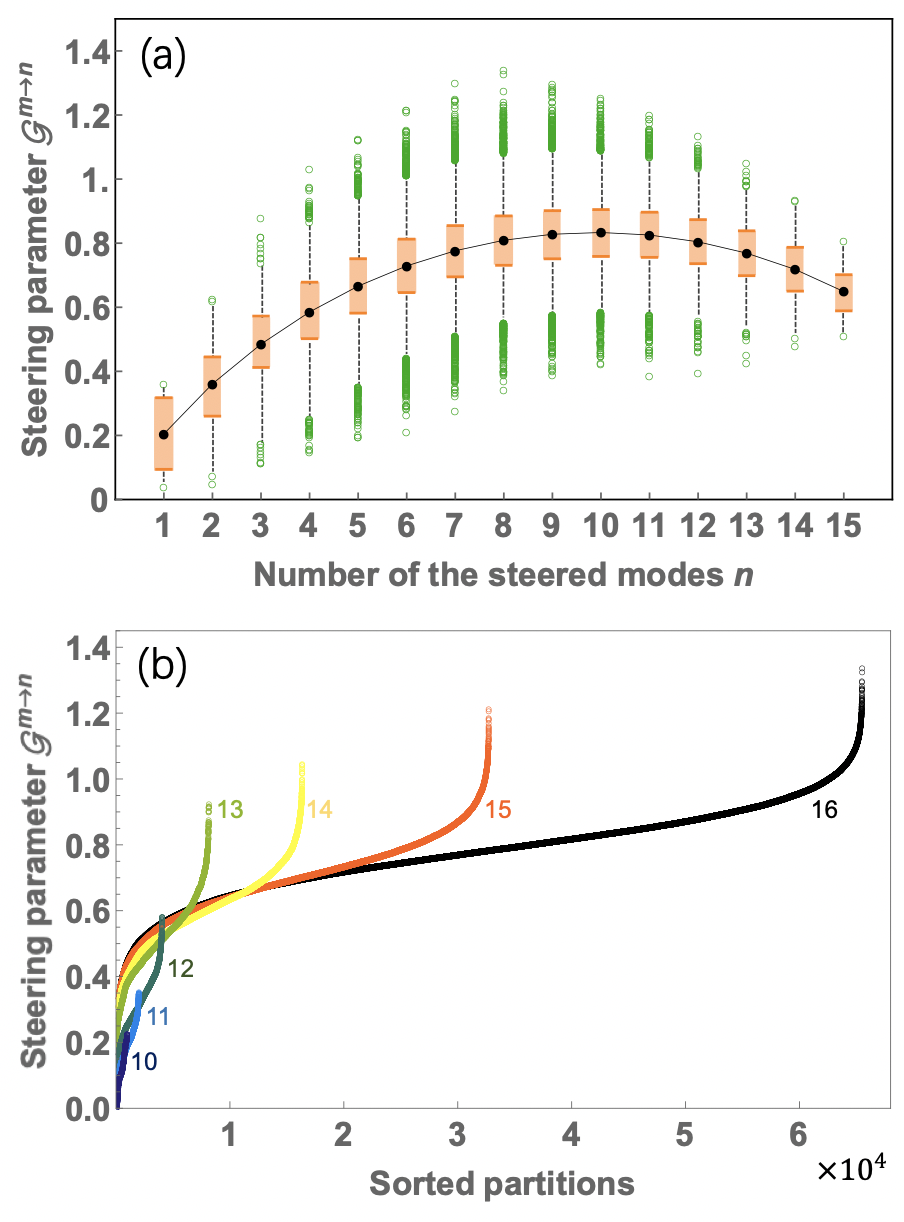}
	\caption{ Multipartite steering among 16 spectral band modes. (a) represents the steerability $\mathcal{G}^{m\to n}$ ($m+n=16$) for all possible 65 534 partitions. The black dots are the average of steerability for each partition. 50\% of data display around the center represented by the orange boxes, and nearly 97\% of data are covered between the whiskers. (b) shows EPR steering against the mode loss of $\{a_{11}\}$, $\{a_{11}, a_{12}\}$, $\{a_{11}, a_{12}, a_{21}\}$, $\ldots$, $\{a_{11}, a_{12}, a_{21}, \ldots, b_{11}, b_{12}\}$, respectively.} %The steerability of the system will continue to reduce more while losing more modes.}
	\label{fig3}
\end{figure}
%

%\blu{3. Quantum steering web and robustness\\
%(1) How many! 65534, Because many modes.\\
%(2) Scalability depends on optical resolution\\
%(3) Robustness. Steering with reduced modes (Loss)}

Thanks to the high spectral resolution of our pulse shaper, EPR steering is finally certified for 65 534 possible bipartitions for the 16-mode system. Fig.~\ref{fig3}(a) illustrates the steerability $\mathcal{G}^{m\to n}$ between $m|n$ bipartition ($m+n=16$). As the prepared state is not a pure state, the trend of steerability is not ideally symmetric, nevertheless the highest steerability still occurs when the steering and steered party are symmetric and cover the full spectrum, as can be seen in the last line of Table~\ref{table1}. The average of steerability is large enough to meet the demand of secure quantum cryptography applications~\cite{Yumonogamy}. 

We also analyze the robustness of multipartite steering against the loss of one or several modes during the transmission, as shown in Fig.~\ref{fig3}(b). We remove up to 6 modes one by one, i.e. removing $\{a_{11}\}$, $\{a_{11}, a_{12}\}$, $\{a_{11}, a_{12}, a_{21}\}$, $\ldots$, $\{a_{11}, a_{12}, a_{21}, a_{22}, b_{11}, b_{12}\}$ from the system, then figure out multipartite steering for all possible partitions of the remaining modes $\mathcal{G}^{m\to n}$ ($m+n=15,14,\ldots,10$). As the information is multimode and spread across the full quantum comb, the system keeps its steerable capability with 32 765, 16 379, 8 185, 4 083, 2 025 and 943 possible nontrivial bipartitions for each reduced systems, respectively. Hence, a fertile and mode-loss resistant repository of the quantum correlations such as multipartite steering and entanglement~\cite{multient2015} is analysed systemically.
%
%\blu{2. Monogamy analysis.}\\
%\blu{(1) Definition four types of monogamy relations.\\
%(2) Demonstrate that four types of monogamy relations remain valid.\\
%(3) Experimental verified that monogamy relations can be break.} 
%
\begin{table}[tb]
	\caption{\label{table2}
		Depiction of four types of monogamy relations and their corresponding experimental cases.}
	\includegraphics[width=1\linewidth]{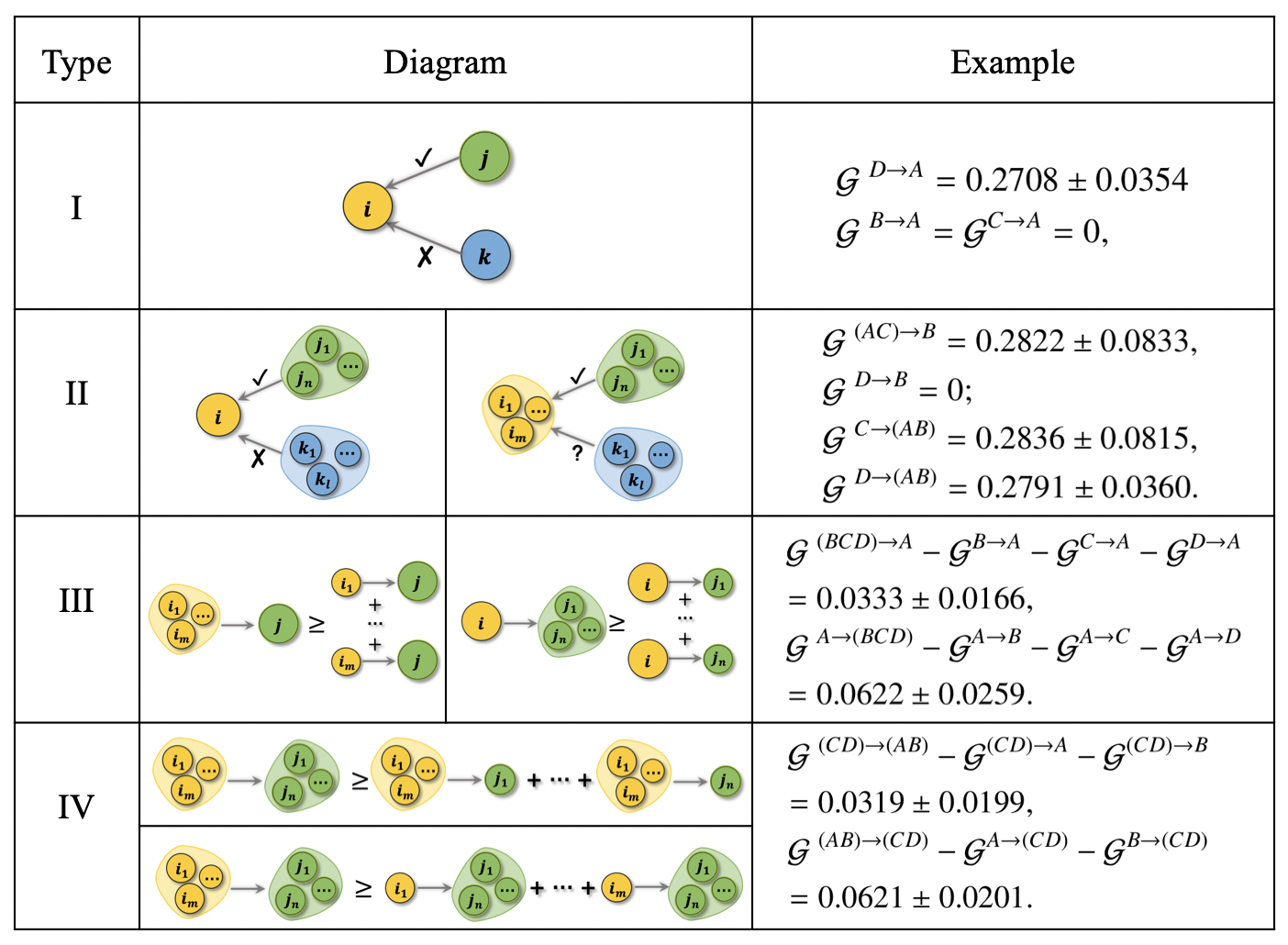}
\end{table}

Finally, to characterize more thoroughly the multimode aspect of the achieved resources, we investigate an important feature called \textit{monogamy relations}, which reveals how EPR steering can be distributed over many different parties. Recently, various kinds of monogamy relations have been studied theoretically~\cite{Reidmonogamy,Kimmonogamy,GSmonogamy, Yumonogamy,Adesso16,G4qubitmonogamy,Shumingmonogamy} and demonstrated experimentally~\cite{prlSu,Shumingmonogamyexp} both for continuous and discrete variable systems. Within the accurate quantification of bipartite Gaussian steering $\mathcal{G}^{m \to n}$, we show that four types of monogamy relations~\cite{Reidmonogamy,Kimmonogamy,GSmonogamy,Yumonogamy,Adesso16} are validated and one of them can be violated in some cases.

As shown in Table \ref{table2}, type-I monogamy relation implies that two distinct modes cannot steer a third mode simultaneously by Gaussian measurements, i.e. Gaussian steering is exclusive~\cite{Reidmonogamy}. Type-II monogamy relation is a generalization of type-I where two steering parties comprise an arbitrary number of modes~\cite{Kimmonogamy,GSmonogamy}. Then, Coffman-Kundu-Wootters (CKW)-type monogamy relation was proposed showing that the sum of steerability between any two modes cannot exceed their intergroup steerability~\cite{Yumonogamy}, which reads $\mathcal{G}^{(i_1,\cdots,i_m) \to j}\geq \mathcal{G}^{i_1 \to j}+\cdots+\mathcal{G}^{i_m \to j}$ and $\mathcal{G}^{i \to (j_1,\cdots,j_n)}\geq \mathcal{G}^{i \to j_1}+\cdots+\mathcal{G}^{i \to j_n}$ ($n,m\geq2$). It was recently extended to a general case (Type-IV) where both steering party and steered party contain more than one mode~\cite{Adesso16}, which takes the form $\mathcal{G}^{(i_1,\cdots,i_m) \to (j_1,\cdots,j_n)}\geq \mathcal{G}^{(i_1,\cdots,i_m) \to j_1}+\cdots+\mathcal{G}^{(i_1,\cdots,i_m) \to j_n}$ and $\mathcal{G}^{(i_1,\cdots,i_m) \to (j_1,\cdots,j_n)}\geq \mathcal{G}^{i_1 \to (j_1,\cdots,j_n)}+\cdots+\mathcal{G}^{i_m \to (j_1,\cdots,j_n)}$ (the latter one can be violated in some cases).

Our experimental results demonstrate that type-I, CKW-type and type-IV remain valid for all possible nontrivial mode partitions regardless of the spectral resolutions. Some of the results are shown in Table \ref{table2}. But type-II monogamy relation can be lifted when the steered party is made of more than one mode, as was predicted theoretically~\cite{GSmonogamy}. Here, we present an experimental observation that  $C\to (AB)$ and $D\to (AB)$ are steerable simultaneously by Gaussian measurements, with $\mathcal{G}^{C\to (AB)}=0.2836\pm0.0815$ and $\mathcal{G}^{D\to (AB)}=0.2791\pm0.0360$ even if the spectrum is divided into only 4 pixels. When the detection spectral resolution increases, the violation will be stronger with more refined spectral band modes, e.g., $\mathcal{G}^{(c_{11}, c_{12},  c_{21}, c_{22}) \to (a_{11}, a_{12}, \ldots, b_{21}, b_{22})}=0.5253\pm0.1117$ and $\mathcal{G}^{(d_{11}, d_{12},  d_{21}, d_{22}) \to (a_{11}, a_{12}, \ldots, b_{21}, b_{22})}=0.4405\pm0.0520$ with 16 pixels. Besides, to obtain the uncertainties in Table \ref{table1} and \ref{table2}, we drew Gaussian-distributed random values with s.d.'s matching those of the experimentally measured quadrature squeezing values \cite{Cainc}. Using these random numbers, numerical steerabilities are obtained by simulating with the use of equation \ref{eq2}.

%\section{Conclusion}
\textit{Conclusion.} In summary, we demonstrate a versatile and flexible repository of multipartite EPR steering within an optical frequency comb and ultrafast pulse shaping. Thanks to this intrinsic multimode resource, we are able to prepare on-demand Gaussian steerable states with 4, 8 and 16 modes through a variable detection spectral resolution without changing the photonics architecture. Increasing the spectral resolution, more independent squeezed modes as well as better squeezing can be extracted and thus the steerability between according bipartition also increases sharply. Then, we identify multipartite steering in all of the 65 534 nontrivial bipartitions of the 16-mode state. As the quantum comb carries all information, the prepared 16-mode state can maintain high steerability in the presence of mode loss. Finally,  we verify four types of monogamy relations of Gaussian steering and demonstrate one possible violation of type-II relation. 

It is theoretically predicted that this quantum comb source can lead to up to a hundred of independent squeezed modes given that one can enhance the energy and spectral bandwidth of the LO light and the spectral resolutions~\cite{100mode}.  In principle, the prepared modes within frequency or spatial pixel could be separated and measured simultaneously with dispersive optics and photodiode arrays~\cite{separatesepctal}. Such a property contributes to the quantum comb being a powerful candidate to construct quantum communication networks in real-world.
%After being separated, frexels could be sent to different parties in a network or directly subject to independent homodyne measurements, for example.

\textbf{Acknowledgments} We acknowledge insightful comments and discussion on multimode entanglement and experiment with C. Fabre and J. Roslund. Y.C. and Y.L. acknowledge the National Key R$\&$D Program of China (Grants No. 2017YFA0303700), the National Natural Science Foundation of China (Grants No. 61975159 and No. 11904279), and the National Science Foundation of Jiangsu Province (Grant No. BK20180322). Y.X. recognizes the National Postdoctoral Program for Innovative Talents (BX20180015), and China Postdoctoral Science Foundation (2019M650291). Q.H. thanks the support from the National Natural Science Foundation of China (Grants No. 11622428 and No. 61675007), the National Key R$\&$D Program of China (Grants No. 2018YFB1107200 and No. 2016YFA0301302), Beijing Natural Science Foundation (Grant No. Z190005), and the Key R$\&$D Program of Guangdong Province (Grant No. 2018B030329001). This work is supported by the French National Research Agency projects COMB and SPOCQ, N.T. acknowledges financial support of the Institut Universitaire de France. 

Y. Cai, Y. Xiang and Y. Liu contributed equally to this work.

\appendix*
\setcounter{equation}{0}
\renewcommand\thefigure{A\arabic{figure}}
\renewcommand\theequation{A\arabic{equation}}
\renewcommand\thetable{A\arabic{table}}
\subsection*{Appendix A: 4-, 8- and 16-mode states of the quantum comb}
\begin{figure}[!htb]
	\centering
	\includegraphics[width=0.85\linewidth]{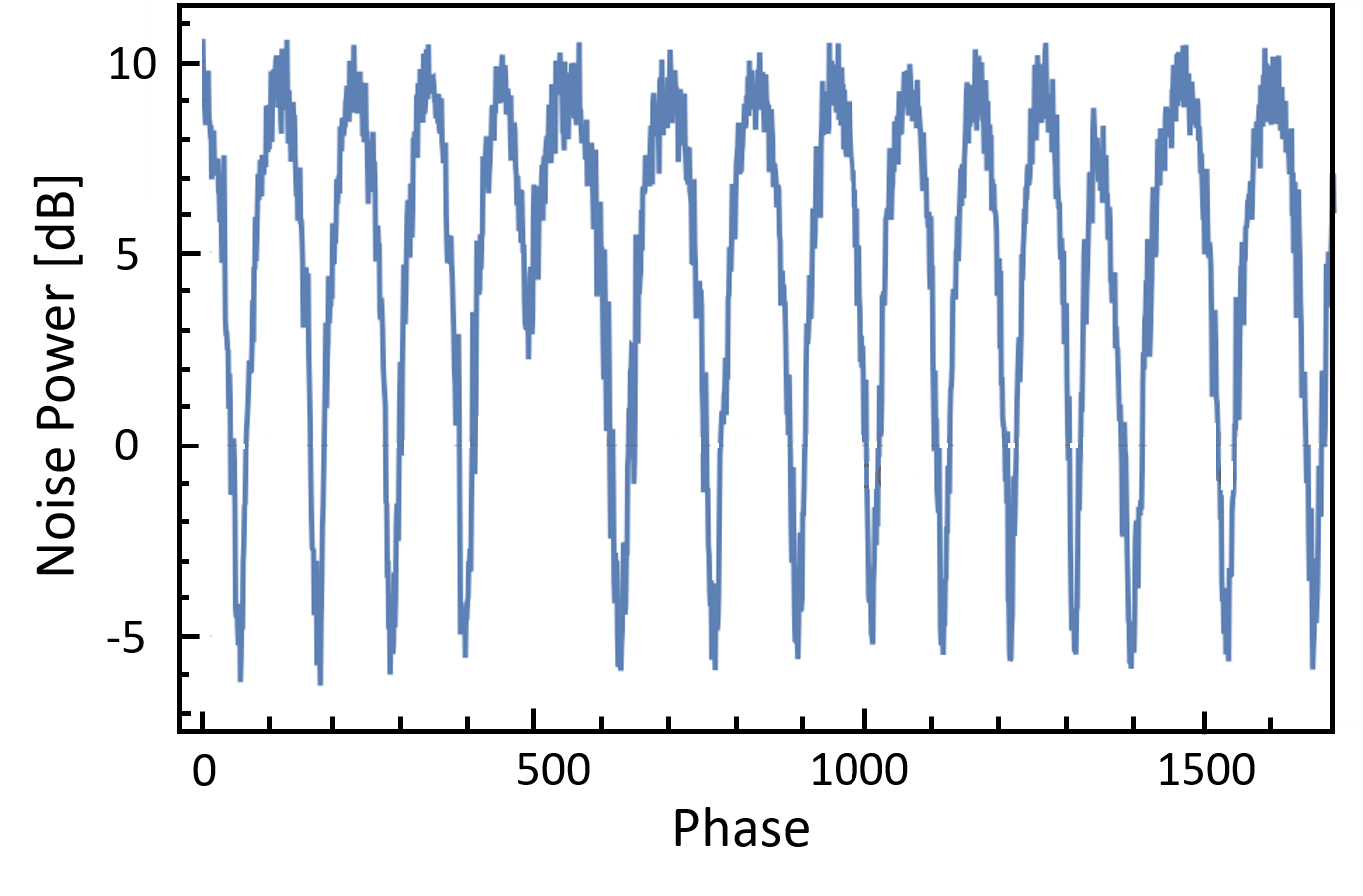}
	\setcounter{figure}{0}
	\caption{Squeezing curve of the quantum comb. The squeezing of quanutum comb is measured via Homodyne detection. Dark noise is corrected. Resolution bandwidth is 100 KHz, and Video bandwidth is 100 Hz,  and 1600 points of variances at 3 MHz are collected in 5 seconds of scan time.} \label{ssfig1}. 
\end{figure}
\begin{figure}[!htb]
	\centering
	\includegraphics[width=\linewidth]{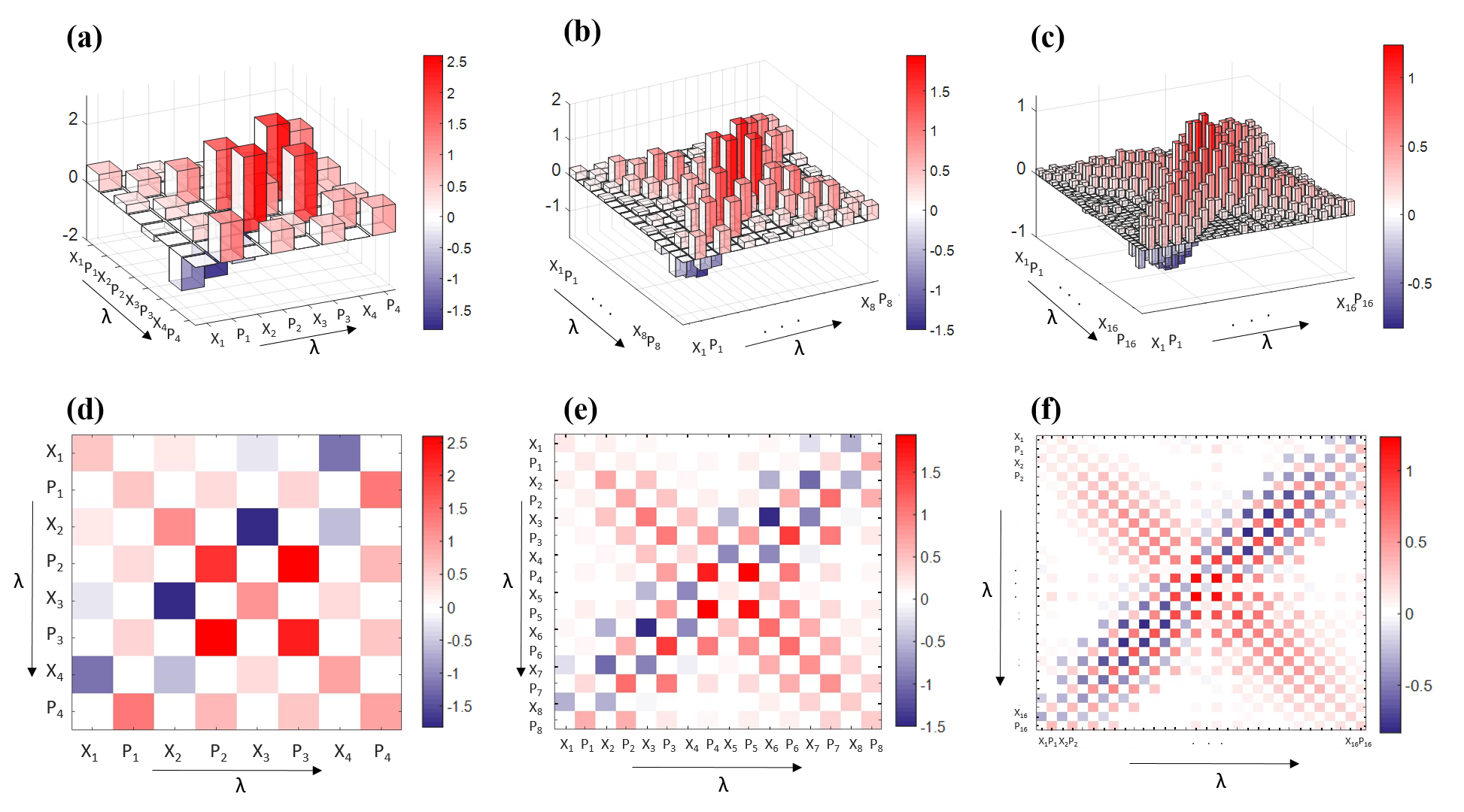}
	\caption{(a)(b)(c) The covariance matrix of 4-, 8- and 16-mode states in the frequency-pixel basis, rexpectively. This correlation matrix is obtained via balanced homodyne detection where the spectrum of the local oscillator is divided into 4, 8 and 16 spectral bands of equivalent width. The shot noise contribution has been subtracted from the diagonal terms for increased visibility. (d)(e)(f) The top views of the corresponding covariance matrix. 15$\%$ loss and dark noise are both corrected.} \label{ssfig2}
\end{figure}

Experimentally, the optical parametric oscillator, synchronously pumped by a train of $\sim100$ fs ultrafast pulses centered at 400 nm with repetition rate 76 MHz, generates about 5 dB squeezing. The measurement curve is shown in Fig. \ref{ssfig1}. Here the visibility of homodyne detection is about 95$\%$, and the quantum efficiency of the photodiode (Hamamastu) is 99$\%$.
The spectrum of local oscillator is divided into 4-, 8-, 16- bands with a variable resolution of the pulses shaping. Quantum correlations in between spectral band modes are interrogated by employing a pulse-shaped homodyne apparatus. As seen in Fig. \ref{ssfig2}, the CMs of 4-, 8- and 16-mode states are reconstructed. The scales of these quantum states can be modulated only by the spectral resolution of the pulse shaping.

\subsection*{Appendix B: Comparison with single mode squeezed states}
\begin{figure}[!htb]
	\centering
	\includegraphics[width=0.8\linewidth]{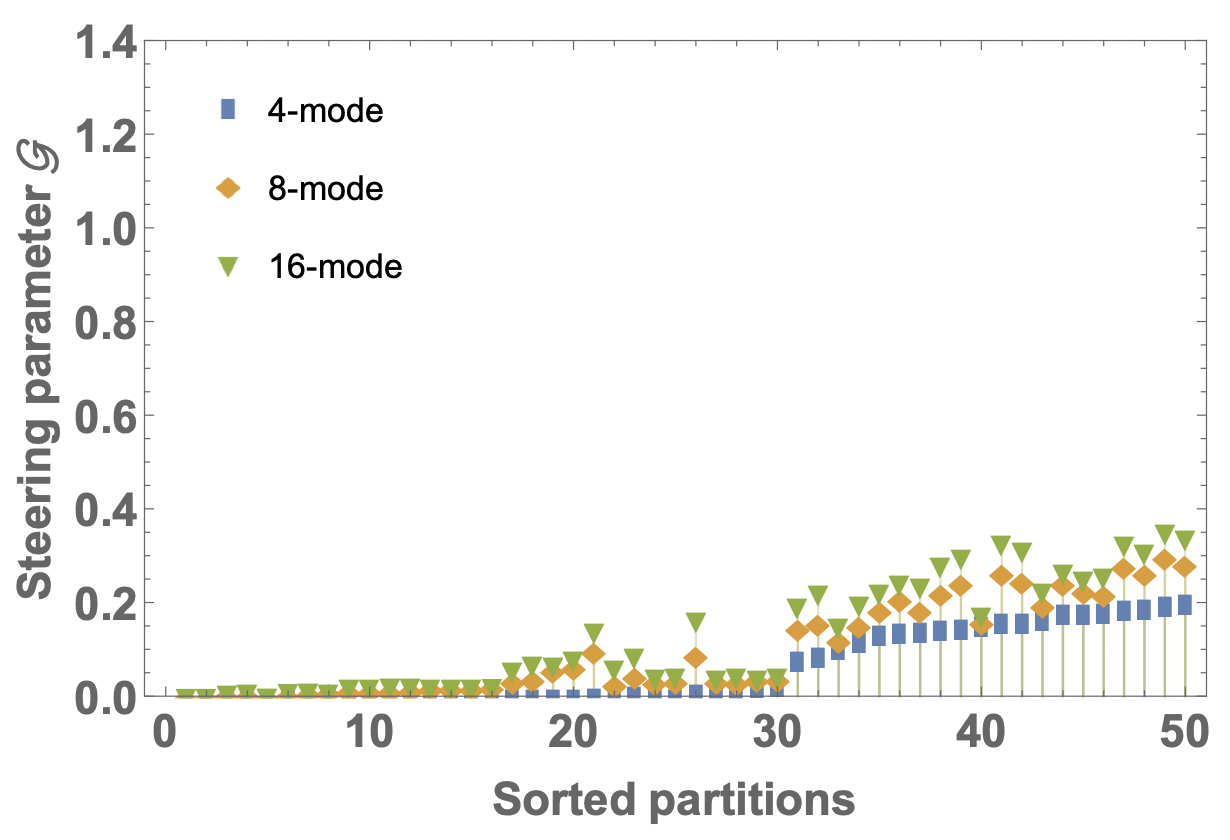}
	\caption{Simulation of EPR steering among the four spectral bands $ABCD$ with a single-eigenmode-squeezed state. The blue rectangle, orange diamond, green down-triangle represent the multipartite EPR steering while the local oscillator is divided into 4, 8, 16 spectral pixels, respectively.} \label{ssfig3}
\end{figure}
In contrast to a multimode squeezed state generated by the quantum comb, we simulate EPR steering using only a single squeezed eigenmode state but the level of squeezing is maintained. As seen in Fig. \ref{ssfig3}, the steerability of 4-, 8-, and 16-mode states are calculated. We find, compared with the Fig. 2 in the main text, for the single squeezed eigenmode case, the steerability almost doesn't change when increasing the spectral resolution from 4 to 16 modes.  

\begin{figure}[b]
	\centering
	\includegraphics[width=0.8\linewidth]{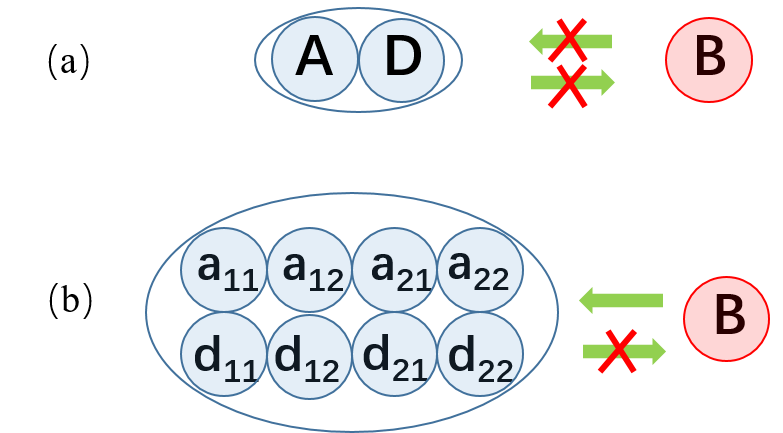}
	\caption{(a) No EPR steering exists in both two directions cross $(AD)-B$ splitting. (b) One-way steering in the direction $B\to(a_{11}, a_{12}, \ldots, d_{21}, d_{22})$ by asymmetically increasing the resolution of one party.} \label{ssfig4}
\end{figure}

\subsection*{Appendix C: Asymmetric control to generate one-way steering}

Different from symmetrical entanglement and Bell nonlocality, one-way steering is a unique phenomenon which means Alice can steer Bob but Bob cannot steer Alice. This feature frequently appears in some asymmetric systems, such as asymmetric photonic systems \cite{OneWayPryde,OneWayGuo} and lossy CV optical networks \cite{OneWayNatPhot,prlSu}. In the main text, we improve the steerability by increasing the spectral resolution of both the steering party and the steered party. Besides, we can realize one-way steering by increasing the resolutions only on one side. 

As shown in Fig. \ref{ssfig4}, when the whole spectrum is divided into 4 pixels, there is no steering with respect to $(AD)-B$ splitting in both two directions. Now when the left side $(AD)$ is subdivided into ($a_{11}, a_{12}, a_{21}, a_{22}, d_{11}, d_{12}, d_{21}, d_{22}$) while the right side remains constant, we can realize one-way steering in the direction $B\to(a_{11}, a_{12}, \ldots, d_{21}, d_{22})$, which reads as $\mathcal{G}^{ B\to(a_{11}, a_{12}, \ldots, d_{21}, d_{22})}=0.1382$ and $\mathcal{G}^{(a_{11}, a_{12}, \ldots, d_{21}, d_{22}) \to B}=0$. 

\subsection*{Appendix D: Data of steerability of 50 partitions with 4-, 8- and 16-pixel resolution}

\begin{center}
	\setcounter{table}{0}
	\tablefirsthead{%
		\hline
		\multicolumn{1}{|c}{Index} &
		\multicolumn{1}{|c|}{Pixels} &
		Partitions &
		\multicolumn{1}{c|}{$\mathcal{G}$} \\
		\hline}
	\tablehead{%
		\hline
		\multicolumn{4}{|l|}{\small\sl continued from previous column}\\
		\hline}
	\tabletail{%
		\hline
		\multicolumn{4}{|r|}{\small\sl continued on next column}\\
		\hline}
	\tablelasttail{\hline}
	\bottomcaption{Steerability of 50 partitions with 4-, 8- and 16-pixel resolution in Fig. 2.}
	\begin{supertabular}{|r@{\hspace{6.5mm}}|r@{\hspace{5.5mm}}|r|r|}
		\multirow{3}[0]{*}{1} & 4     & $\mathcal{G}^{B\to A}$   & 0 \\
		& 8     & $\mathcal{G}^{(b_1,b_2) \to (a_1,a_2)}$ & 0 \\
		& 16    & $\mathcal{G}^{(b_{11},b_{12},b_{21},b_{22}) \to (a_{11},a_{12},a_{21},a_{22})}$ & 0 \\
		\hline
		\multirow{3}[0]{*}{2} & 4     & $\mathcal{G}^{D\to B}$& 0  \\
		& 8     & $\mathcal{G}^{(d_1,d_2) \to (b_1,b_2)}$ & 0  \\
		& 16    & $\mathcal{G}^{(d_{11},d_{12},d_{21},d_{22}) \to (b_{11},b_{12},b_{21},b_{22})}$ & 0  \\
		\hline
		\multirow{3}[0]{*}{3} & 4     & $\mathcal{G}^{C\to A}$& 0  \\
		& 8     & $\mathcal{G}^{(c_1,c_2) \to (a_1,a_2)}$ & 0 \\
		& 16    & $\mathcal{G}^{(c_{11},c_{12},c_{21},c_{22}) \to (a_{11},a_{12},a_{21},a_{22})}$& 0  \\
		\hline
		\multirow{3}[0]{*}{4} & 4     & $\mathcal{G}^{D\to C}$ & 0 \\
		& 8     &  $\mathcal{G}^{(d_1,d_2) \to (c_1,c_2)}$ & 0  \\
		& 16    &  $\mathcal{G}^{(d_{11},d_{12},d_{21},d_{22}) \to (c_{11},c_{12},c_{21},c_{22})}$ & 0\\
		\hline
		\multirow{3}[0]{*}{5} & 4     & $\mathcal{G}^{C\to D}$ & 0 \\
		& 8     &  $\mathcal{G}^{(c_1,c_2) \to (d_1,d_2)}$ & 0  \\
		& 16    &  $\mathcal{G}^{ (c_{11},c_{12},c_{21},c_{22})\to (d_{11},d_{12},d_{21},d_{22})}$ & 0\\
		\hline
		\multirow{3}[0]{*}{6} & 4     & $\mathcal{G}^{(A,D)\to C}$& 0 \\
		& 8     & $\mathcal{G}^{(a_1,a_2,d_1,d_2) \to (c_1,c_2)}$  & 0 \\
		& 16    & $\mathcal{G}^{ (a_{11},a_{12},...,d_{21},d_{22})\to (c_{11},c_{12},c_{21},c_{22})}$& 0.0039  \\
		\hline
		\multirow{3}[0]{*}{7} & 4     &  $\mathcal{G}^{(A,D)\to B}$ & 0\\
		& 8     & $\mathcal{G}^{(a_1,a_2,d_1,d_2) \to (b_1,b_2)}$& 0 \\
		& 16    & $\mathcal{G}^{ (a_{11},a_{12},...,d_{21},d_{22})\to (b_{11},b_{12},b_{21},b_{22})}$& 0.0046  \\
		\hline
		\multirow{3}[0]{*}{8} & 4     &  $\mathcal{G}^{D\to B}$ & 0 \\
		& 8     &  $\mathcal{G}^{(d_1,d_2) \to (b_1,b_2)}$& 0 \\
		& 16    &$\mathcal{G}^{(d_{11},d_{12},d_{21},d_{22}) \to (b_{11},b_{12},b_{21},b_{22})}$& 0.0140  \\
		\hline
		\multirow{3}[0]{*}{9} & 4     & $\mathcal{G}^{(B,C)\to D}$ & 0  \\
		& 8     &$\mathcal{G}^{(b_1,b_2,c_1,c_2) \to (d_1,d_2)}$ & 0 \\
		& 16    & $\mathcal{G}^{ (b_{11},b_{12},...,c_{21},c_{22})\to (d_{11},d_{12},d_{21},d_{22})}$ & 0.0161  \\
		\hline
		\multirow{3}[0]{*}{10} & 4     & $\mathcal{G}^{B\to A}$ & 0  \\
		& 8     & $\mathcal{G}^{(b_1,b_2) \to (a_1,a_2)}$& 0  \\
		& 16    &$\mathcal{G}^{(b_{11},b_{12},b_{21},b_{22}) \to (a_{11},a_{12},a_{21},a_{22})}$ & 0.0214  \\
		\hline
		\multirow{3}[0]{*}{11} & 4     & $\mathcal{G}^{C\to A}$& 0  \\
		& 8     & $\mathcal{G}^{(c_1,c_2) \to (a_1,a_2)}$ & 0  \\
		& 16    & $\mathcal{G}^{(c_{11},c_{12},c_{21},c_{22}) \to (a_{11},a_{12},a_{21},a_{22})}$ & 0.0547  \\
		\hline
		\multirow{3}[0]{*}{12} & 4     & $\mathcal{G}^{(B,C)\to A}$ & 0  \\
		& 8     & $\mathcal{G}^{(b_1,b_2,c_1,c_2) \to (a_1,a_2)}$ & 0  \\
		& 16    & $\mathcal{G}^{(b_{11},b_{12},...,c_{21},c_{22}) \to (a_{11},a_{12},a_{21},a_{22})}$ & 0.0644  \\
		\hline
		\multirow{3}[0]{*}{13} & 4     & $\mathcal{G}^{A \to (B,C)}$ & 0 \\
		& 8     & $\mathcal{G}^{(a_1,a_2) \to (b_1,b_2,c_1,c_2)}$ & 0  \\
		& 16    &  $\mathcal{G}^{ (a_{11},a_{12},a_{21},a_{22})\to (b_{11},b_{12},...,c_{21},c_{22})}$ & 0.1481  \\
		\hline
		\multirow{3}[0]{*}{14} & 4     & $\mathcal{G}^{C \to (A,D)}$& 0  \\
		& 8     & $\mathcal{G}^{(c_1,c_2) \to (a_1,a_2,d_1,d_2)}$& 0.0225  \\
		& 16    & $\mathcal{G}^{ (c_{11},c_{12},c_{21},c_{22})\to (a_{11},a_{12},...,d_{21},d_{22})}$& 0.1712  \\
		\hline
		\multirow{3}[0]{*}{15} & 4     & $\mathcal{G}^{D \to (B,C)}$& 0 \\
		& 8     &$\mathcal{G}^{(d_1,d_2) \to (b_1,b_2,c_1,c_2)}$& 0.0250  \\
		& 16    & $\mathcal{G}^{ (d_{11},d_{12},d_{21},d_{22})\to (b_{11},b_{12},...,c_{21},c_{22})}$ & 0.1741  \\
		\hline
		\multirow{3}[0]{*}{16} & 4     &$\mathcal{G}^{B \to (A,D)}$ & 0 \\
		& 8     &  $\mathcal{G}^{(b_1,b_2) \to (a_1,a_2,d_1,d_2)}$& 0.0468  \\
		& 16    & $\mathcal{G}^{ (b_{11},b_{12},b_{21},b_{22})\to (a_{11},a_{12},...,d_{21},d_{22})}$ & 0.2122  \\
		\hline
		\multirow{3}[0]{*}{17} & 4     & $\mathcal{G}^{(A,D) \to (B,C)}$ & 0\\
		& 8     & $\mathcal{G}^{(a_1,a_2,d_1,d_2) \to (b_1,b_2,c_1,c_2)}$ & 0.1171  \\
		& 16    &  $\mathcal{G}^{ (a_{11},a_{12},...,d_{21},d_{22})\to (b_{11},b_{12},...,c_{21},c_{22})}$ & 0.3922 \\
		\hline
		\multirow{3}[0]{*}{18} & 4     & $\mathcal{G}^{(B,C) \to (A,D)}$ & 0.0199  \\
		& 8     & $\mathcal{G}^{ (b_1,b_2,c_1,c_2)\to(a_1,a_2,d_1,d_2) }$& 0.1377  \\
		& 16    & $\mathcal{G}^{ (b_{11},b_{12},...,c_{21},c_{22})\to(a_{11},a_{12},...,d_{21},d_{22}) }$ & 0.4112
		\\
		\hline
		\multirow{3}[0]{*}{19} & 4     &  $\mathcal{G}^{A\to D}$ & 0.0652  \\
		& 8     & $\mathcal{G}^{ (a_1,a_2)\to(d_1,d_2) }$& 0.1322  \\
		& 16    &  $\mathcal{G}^{ (a_{11},a_{12},a_{21},a_{22})\to(d_{11},d_{12},d_{21},d_{22}) }$ & 0.1464  \\
		\hline
		\multirow{3}[0]{*}{20} & 4     & $\mathcal{G}^{(A,C)\to D}$& 0.0859  \\
		& 8     & $\mathcal{G}^{ (a_1,a_2,c_1,c_2)\to(d_1,d_2) }$ & 0.1442  \\
		& 16    &$\mathcal{G}^{ (a_{11},a_{12},...,c_{21},c_{22})\to(d_{11},d_{12},d_{21},d_{22}) }$& 0.1744  \\
		\hline
		\multirow{3}[0]{*}{21} & 4     & $\mathcal{G}^{A\to (C,D)}$ & 0.0972  \\
		& 8     & $\mathcal{G}^{ (a_1,a_2)\to(c_1,c_2,d_1,d_2) }$ & 0.1451  \\
		& 16    & $\mathcal{G}^{ (a_{11},a_{12},a_{21},a_{22})\to(c_{11},c_{12},...,d_{21},d_{22}) }$& 0.2437  \\
		\hline
		\multirow{3}[0]{*}{22} & 4     & $\mathcal{G}^{A\to (B,D)}$ & 0.1158  \\
		& 8     & $\mathcal{G}^{ (a_1,a_2)\to(b_1,b_2,d_1,d_2) }$& 0.2026  \\
		& 16    & $\mathcal{G}^{ (a_{11},a_{12},a_{21},a_{22})\to(b_{11},b_{12},...,d_{21},d_{22}) }$& 0.2854  \\
		\hline
		\multirow{3}[0]{*}{23} & 4     &$\mathcal{G}^{A\to (B,C,D)}$   & 0.1274  \\
		& 8     & $\mathcal{G}^{ (a_1,a_2)\to(b_1,b_2,c_1,c_2,d_1,d_2) }$  & 0.3035  \\
		& 16    & $\mathcal{G}^{ (a_{11},a_{12},a_{21},a_{22})\to(b_{11},b_{12},...,c_{11},...,d_{22}) }$  & 0.6586  \\
		\hline
		\multirow{3}[0]{*}{24} & 4     &$\mathcal{G}^{(A,B)\to D}$ & 0.1335  \\
		& 8     & $\mathcal{G}^{ (a_1,a_2,b_1,b_2)\to(d_1,d_2) }$& 0.2459  \\
		& 16    & $\mathcal{G}^{ (a_{11},a_{12},...,b_{21},b_{22})\to(d_{11},d_{12},d_{21},d_{22}) }$& 0.3552  \\
		\hline
		\multirow{3}[0]{*}{25} & 4     &  $\mathcal{G}^{(A,B,C)\to D}$  & 0.1540  \\
		& 8     &$\mathcal{G}^{(a_1,a_2,b_1,b_2,c_1,c_2) \to (d_1,d_2)}$   & 0.3542  \\
		& 16    &  $\mathcal{G}^{ (a_{11},a_{12},a_{21},a_{22})\to(b_{11},b_{12},...,d_{21},d_{22}) }$  & 0.5241  \\
		\hline
		\multirow{3}[0]{*}{26} & 4     & $\mathcal{G}^{B\to C}$ & 0.2540  \\
		& 8     & $\mathcal{G}^{(b_1,b_2) \to (c_1,c_2)}$& 0.3031  \\
		& 16    &  $\mathcal{G}^{ (b_{11},b_{12},b_{21},b_{22})\to(c_{11},c_{12},c_{21},c_{22}) }$ & 0.4135  \\
		\hline
		\multirow{3}[0]{*}{27} & 4     &  $\mathcal{G}^{(B,D)\to C}$& 0.2567  \\
		& 8     & $\mathcal{G}^{(b_1,b_2,d_1,d_2) \to (c_1,c_2)}$ & 0.3185  \\
		& 16    &  $\mathcal{G}^{ (b_{11},b_{12},...,d_{21},d_{22})\to(c_{11},c_{12},c_{21},c_{22}) }$ & 0.4407  \\
		\hline
		\multirow{3}[0]{*}{28} & 4     &  $\mathcal{G}^{B\to (C,D)}$& 0.2580  \\
		& 8     &  $\mathcal{G}^{(b_1,b_2) \to (c_1,c_2,d_1,d_2)}$& 0.3417  \\
		& 16    &  $\mathcal{G}^{ (b_{11},b_{12},b_{21},b_{22})\to(c_{11},c_{12},...,d_{21},d_{22}) }$ & 0.5048  \\
		\hline
		\multirow{3}[0]{*}{29} & 4     &$\mathcal{G}^{(A,B)\to C}$ & 0.2633  \\
		& 8     & $\mathcal{G}^{(a_1,a_2,b_1,b_2) \to (c_1,c_2)}$& 0.3537  \\
		& 16    &  $\mathcal{G}^{ (a_{11},a_{12},...,b_{21},b_{22})\to(c_{11},c_{12},c_{21},c_{22}) }$ & 0.5177  \\
		\hline
		\multirow{3}[0]{*}{30} & 4     &$\mathcal{G}^{B\to (A,C)}$& 0.2686  \\
		& 8     & $\mathcal{G}^{(b_1,b_2) \to (a_1,a_2,c_1,c_2)}$ & 0.3539  \\
		& 16    &  $\mathcal{G}^{ (b_{11},b_{12},b_{21},b_{22})\to(a_{11},a_{12},...,c_{21},c_{22}) }$& 0.5421  \\
		\hline
		\multirow{3}[0]{*}{31} & 4     & $\mathcal{G}^{D\to A}$ & 0.2708  \\
		& 8     & $\mathcal{G}^{(d_1,d_2) \to (a_1,a_2)}$ & 0.2765  \\
		& 16    &  $\mathcal{G}^{ (d_{11},d_{12},d_{21},d_{22})\to(a_{11},a_{12},a_{21},a_{22}) }$ & 0.3423  \\
		\hline
		\multirow{3}[0]{*}{32} & 4     &$\mathcal{G}^{(C,D)\to A}$ & 0.2723  \\
		& 8     & $\mathcal{G}^{(c_1,c_2,d_1,d_2) \to (a_1,a_2)}$ & 0.2918  \\
		& 16    &  $\mathcal{G}^{ (c_{11},c_{12},...,d_{21},d_{22})\to(a_{11},a_{12},a_{21},a_{22}) }$ & 0.4159  \\
		\hline
		\multirow{3}[0]{*}{33} & 4     & $\mathcal{G}^{D\to (A,C)}$ & 0.2727  \\
		& 8     &  $\mathcal{G}^{(d_1,d_2) \to (a_1,a_2,c_1,c_2)}$& 0.2922  \\
		& 16    &  $\mathcal{G}^{ (d_{11},d_{12},d_{21},d_{22})\to(a_{11},a_{12},...,c_{21},c_{22}) }$ & 0.4051  \\
		\hline
		\multirow{3}[0]{*}{34} & 4     & $\mathcal{G}^{C\to B}$ & 0.2768  \\
		& 8     & $\mathcal{G}^{(c_1,c_2) \to (b_1,b_2)}$ & 0.3239  \\
		& 16    & $\mathcal{G}^{ (c_{11},c_{12},c_{21},c_{22})\to(b_{11},b_{12},b_{21},b_{22}) }$ & 0.4021  \\
		\hline
		\multirow{3}[0]{*}{35} & 4     & $\mathcal{G}^{(B,D)\to A}$& 0.2770  \\
		& 8     & $\mathcal{G}^{(b_1,b_2,d_1,d_2) \to (a_1,a_2)}$ & 0.2970  \\
		& 16    & $\mathcal{G}^{ (b_{11},b_{12},...,d_{21},d_{22})\to(a_{11},a_{12},a_{21},a_{22}) }$  & 0.3860  \\
		\hline
		\multirow{3}[0]{*}{36} & 4     & $\mathcal{G}^{D\to (A,B)}$ & 0.2791  \\
		& 8     & $\mathcal{G}^{(d_1,d_2) \to (a_1,a_2,b_1,b_2)}$ & 0.3132  \\
		& 16    & $\mathcal{G}^{ (d_{11},d_{12},d_{21},d_{22})\to(a_{11},a_{12},...,b_{21},b_{22}) }$ & 0.4405  \\
		\hline
		\multirow{3}[0]{*}{37} & 4     & $\mathcal{G}^{B\to (A,C)}$& 0.2822  \\
		& 8     &  $\mathcal{G}^{ (b_1,b_2)\to (a_1,a_2,c_1,c_2)}$& 0.3362  \\
		& 16    & $\mathcal{G}^{(b_{11},b_{12},b_{21},b_{22})\to (a_{11},a_{12},...,c_{21},c_{22}) }$  & 0.4318  \\
		\hline
		\multirow{3}[0]{*}{38} & 4     & $\mathcal{G}^{C\to (A,B)}$& 0.2836  \\
		& 8     & $\mathcal{G}^{(c_1,c_2) \to (a_1,a_2,b_1,b_2)}$& 0.3446  \\
		& 16    & $\mathcal{G}^{ (c_{11},c_{12},c_{21},c_{22})\to(a_{11},a_{12},...,b_{21},b_{22}) }$ & 0.5253  \\
		\hline
		\multirow{3}[0]{*}{39} & 4     &  $\mathcal{G}^{(A,B,D)\to C}$& 0.2837  \\
		& 8     &   $\mathcal{G}^{(a_1,a_2,b_1,b_2,d_1,d_2) \to (c_1,c_2)}$ & 0.4529  \\
		& 16    &$\mathcal{G}^{ (a_{11},a_{12},...,b_{12},...,d_{22})\to(c_{11},c_{12},c_{21},c_{22}) }$    & 0.6886  \\
		\hline
		\multirow{3}[0]{*}{40} & 4     &  $\mathcal{G}^{(B,C,D)\to A}$ & 0.3041  \\
		& 8     &$\mathcal{G}^{(b_1,b_2,c_1,c_2,d_1,d_2) \to (a_1,a_2)}$  & 0.3955  \\
		& 16    & $\mathcal{G}^{ (b_{11},b_{12},...,c_{12},...,d_{22})\to(a_{11},a_{12},a_{21},a_{22}) }$   & 0.6158  \\
		\hline
		\multirow{3}[0]{*}{41} & 4     & $\mathcal{G}^{C\to (B,D)}$& 0.3065  \\
		& 8     & $\mathcal{G}^{(c_1,c_2) \to (b_1,b_2,d_1,d_2)}$& 0.3940  \\
		& 16    &$\mathcal{G}^{ (c_{11},c_{12},c_{21},c_{22})\to(b_{11},b_{12},...,d_{21},d_{22}) }$  & 0.5229  \\
		\hline
		\multirow{3}[0]{*}{42} & 4     & $\mathcal{G}^{(C,D)\to B}$ & 0.3184  \\
		& 8     & $\mathcal{G}^{(c_1,c_2,d_1,d_2) \to (b_1,b_2)}$ & 0.4533  \\
		& 16    & $\mathcal{G}^{ (c_{11},c_{12},...,d_{21},d_{22})\to(b_{11},b_{12},b_{21},b_{22}) }$ & 0.5825  \\
		\hline
		\multirow{3}[0]{*}{43} & 4     &  $\mathcal{G}^{D\to (A,B,C)}$ & 0.3229  \\
		& 8     & $\mathcal{G}^{(d_1,d_2) \to (a_1,a_2,b_1,b_2,c_1,c_2)}$& 0.5681  \\
		& 16    &$\mathcal{G}^{ (d_{11},d_{12},d_{21},d_{22})\to(a_{11},a_{12},...,b_{12},...,c_{22}) }$   & 0.9428  \\
		\hline
		\multirow{3}[0]{*}{44} & 4     &    $\mathcal{G}^{B\to (A,C,D)}$& 0.3260  \\
		& 8     &   $\mathcal{G}^{(b_1,b_2) \to (a_1,a_2,c_1,c_2,d_1,d_2)}$ & 0.5681  \\
		& 16    & $\mathcal{G}^{ (b_{11},b_{12},b_{21},b_{22})\to(a_{11},a_{12},...,c_{12},...,d_{22}) }$ & 0.9643  \\
		\hline
		\multirow{3}[0]{*}{45} & 4     & $\mathcal{G}^{C\to (A,B,D)}$ & 0.3386  \\
		& 8     &  $\mathcal{G}^{(c_1,c_2) \to (a_1,a_2,b_1,b_2,d_1,d_2)}$  & 0.6210  \\
		& 16    & $\mathcal{G}^{ (c_{11},c_{12},c_{21},c_{22})\to(a_{11},a_{12},...,b_{12},...,d_{22}) }$  & 1.0056  \\
		\hline
		\multirow{3}[0]{*}{46} & 4     & $\mathcal{G}^{(A,B,D)\to C}$& 0.3503  \\
		& 8     & $\mathcal{G}^{(a_1,a_2,b_1,b_2,d_1,d_2) \to (c_1,c_2) }$  & 0.5995  \\
		& 16    &$\mathcal{G}^{ (a_{11},a_{12},...,b_{12},...,d_{22})\to(c_{11},c_{12},c_{21},c_{22}) }$ & 0.8097  \\
		\hline
		\multirow{3}[0]{*}{47} & 4     &  $\mathcal{G}^{(A,B)\to (C,D)}$& 0.4173  \\
		& 8     & $\mathcal{G}^{(a_1,a_2,b_1,b_2) \to (c_1,c_2,d_1,d_2) }$& 0.7865  \\
		& 16    &$\mathcal{G}^{ (a_{11},a_{12},...,b_{21},b_{22})\to(c_{11},c_{12},...,d_{21},d_{22}) }$ & 1.1667  \\
		\hline
		\multirow{3}[0]{*}{48} & 4     &  $\mathcal{G}^{(A,C)\to (B,D)}$& 0.4362  \\
		& 8     & $\mathcal{G}^{(a_1,a_2,c_1,c_2) \to (b_1,b_2,d_1,d_2) }$ & 0.7219  \\
		& 16    & $\mathcal{G}^{ (a_{11},a_{12},...,c_{21},c_{22})\to(b_{11},b_{12},...,d_{21},d_{22}) }$ & 0.9750  \\
		\hline
		\multirow{3}[0]{*}{49} & 4     & $\mathcal{G}^{(B,D)\to (A,C)}$ & 0.5608  \\
		& 8     & $\mathcal{G}^{(b_1,b_2,d_1,d_2) \to (a_1,a_2,c_1,c_2) }$& 0.7714  \\
		& 16    &$\mathcal{G}^{ (b_{11},b_{12},...,d_{21},d_{22})\to(a_{11},a_{12},...,c_{21},c_{22}) }$ & 1.0906  \\
		\hline
		\multirow{3}[0]{*}{50} & 4     & $\mathcal{G}^{(C,D)\to (A,B)}$ & 0.6226  \\
		& 8     & $\mathcal{G}^{(c_1,c_2,d_1,d_2) \to (a_1,a_2,b_1,b_2) }$ & 0.9692  \\
		& 16    & $\mathcal{G}^{(c_{11},c_{12},...,d_{21},d_{22})\to(a_{11},a_{12},...,b_{21},b_{22}) }$& 1.3415  \\
	\end{supertabular}
\end{center}

%%%%%%%%%%%%%%%%%%%%%%%%%%%%%%%%%%%


\begin{thebibliography}{99}
	
\bibitem{EPR35} A. Einstein, B. Podolsky, and N. Rosen, Can quantum-mechanical description of physical reality be considered complete? Phys. Rev. \textbf{47}, 777--780 (1935).
	
\bibitem{Schrodinger35} E. Schr\"{o}dinger, Discussion of probability
relations between separated systems, Proc. Cambridge Philos. Soc. \textbf{31%
}, 555--563 (1935).

\bibitem{Howard07PRL} H. M. Wiseman, S. J. Jones, and A. C. Doherty,
Steering, entanglement, nonlocality, and the Einstein-Podolsky-Rosen
paradox, Phys. Rev. Lett. \textbf{98}, 140402 (2007).

\bibitem{Howard07PRA} S. J. Jones, H. M. Wiseman, and A. C. Doherty,
Entanglement, Einstein-Podolsky-Rosen correlations, Bell nonlocality, and
steering, Phys. Rev. A \textbf{76}, 052116 (2007).

\bibitem{Eric13} E. G. Cavalcanti, M. J. W. Hall, and H. M. Wiseman, Entanglement verification and steering when Alice and Bob cannot be trusted, Phys. Rev. A \textbf{87}, 032306 (2013).

\bibitem{cavalcanti17review} D. Cavalcanti and P. Skrzypczyk, Quantum
steering: a review with focus on semidefinite programming, Rep. Prog. Phys.
\textbf{80}, 024001 (2017).

\bibitem{ReidRMP} M. D. Reid, P. D. Drummond, W. P. Bowen, E. G. Cavalcanti, P. K. Lam, H. A. Bachor, U. L. Andersen, and G. Leuchs, Colloquium: The Einstein-Podolsky-Rosen paradox: From concepts to applications, Rev. Mod. Phys. \textbf{81}, 1727--1751 (2009). Roope Uola, Ana C. S. Costa, H. Chau Nguyen, Otfried G$\ddot{u}$hne, Quantum steering, arXiv \textbf{1903}, 06663(2019).


\bibitem{BrunnerRMP} N. Brunner, D. Cavalcanti, S. Pironio, V. Scarani, and
S. Wehner, Bell nonlocality, Rev. Mod. Phys. \textbf{86}, 419--478 (2014).

\bibitem{entRMP} R. Horodecki, P. Horodecki, M. Horodecki, and K. Horodecki,
Quantum entanglement, Rev. Mod. Phys. \textbf{81}, 865--942 (2009).




\bibitem{one-way-Theory} S. L.W. Midgley, A. J. Ferris, and M. K. Olsen, Asymmetric
Gaussian steering: When Alice and Bob disagree, Phys. Rev. A \textbf{81}, 022101 (2010); S. P.Walborn, A. Salles, R. M. Gomes, F. Toscano, and P. H. Souto Ribeiro, Revealing hidden Einstein-Podolsky-Rosen nonlocality, Phys. Rev. Lett. \textbf{106}, 130402 (2011); J. Schneeloch, C. J. Broadbent, S. P. Walborn, E. G.
Cavalcanti, and J. C. Howell, Einstein-Podolsky-Rosen steering inequalities from entropic uncertainty relations, Phys. Rev. A \textbf{87}, 062103 (2013); J.
Bowles, T.Vertesi, M. T. Quintino, and N. Brunner, One-way Einstein-Podolsky-Rosen steering, Phys. Rev. Lett. \textbf{112}, 200402 (2014); B. Opanchuk, L. Arnaud, and M. D. Reid, Detecting faked continuous-variable entanglement using one-sided
device-independent entanglement witnesses, Phys. Rev. A \textbf{89}, 062101 (2014).

\bibitem{He15} Q. Y. He, Q. H. Gong, and M. D. Reid, Classifying
directional Gaussian entanglement, Einstein-Podolsky-Rosen steering, and
discord, Phys. Rev. Lett. \textbf{114}, 060402 (2015).

\bibitem{ReidJOSAB} L. Rosales-Z\'{a}rate, R. Y. Teh, S. Kiesewetter, A.
Brolis, K. Ng, and M. D. Reid, Decoherence of
Einstein-Podolsky-Rosen steering, J. Opt. Soc. Am. B \textbf{32}, A82-A91
(2015).

\bibitem{SQT13Reid} M. D. Reid. Signifying quantum benchmarks for qubit teleportation and secure quantum communication using Einstein-Podolsky-Rosen steering inequalities, Phys. Rev. A, \textbf{88}, 062338 (2013).

\bibitem{SQT15} Q. He, L. Rosales-Z\'{a}rate, G. Adesso, and M. D. Reid, Secure continuous variable teleportation and Einstein-Podolsky-Rosen steering, Phys. Rev. Lett. \textbf{115}, 180502 (2015).

\bibitem{SQT16_LiCM} C.-Y. Chiu, N. Lambert, Teh-Lu Liao, F. Nori, and C.-M. Li, No-cloning of quantum steering, NPJ Quantum Information \textbf{2}, 16020 (2016).

\bibitem{1sDIQKD} M. Tomamichel and R. Renner, Uncertainty relation for smooth entropies, Phys. Rev. Lett. \textbf{106}, 110506 (2011).

\bibitem{1sDIQKD_howard} C. Branciard, E. G. Cavalcanti, S. P. Walborn, V. Scarani, and H. M. Wiseman, One-sided device-independent quantum key distribution: security, feasibility, and the connection with steering, Phys. Rev. A \textbf{85}, 010301 (2012).

\bibitem{CV-QKDexp} T. Gehring, V. H\"{a}ndchen, J. Duhme, F. Furrer, T. Franz, C. Pacher, R. F. Werner, and R. Schnabel, Implementation of continuous-variable quantum key distribution with composable and one-sided-device independent security against coherent attacks, Nat. Commun. \textbf{6}, 8795 (2015).

\bibitem{HowardOptica} N. Walk, S. Hosseini, J. Geng, O. Thearle, J. Y. Haw, S. Armstrong, S. M. Assad, J. Janou\v{s}ek, T. C. Ralph, T. Symul, H. M. Wiseman, and P. K. Lam, Experimental demonstration of Gaussian protocols for one-sided device-independent quantum key distribution, Optica \textbf{3}, 634--642 (2016).

\bibitem{subchannel} M. Piani and J. Watrous, Necessary and sufficient quantum information characterization of Einstein-Podolsky-Rosen steering, Phys. Rev. Lett. \textbf{114}, 060404 (2015).

\bibitem{subchannel16} S.-L. Chen, C. Budroni, Y.-C. Liang, and Y.-N. Chen, Natural framework for device-independent quantification of quantum steerability, measurement incompatibility, and self-testing, Phys. Rev. Lett. \textbf{116}, 240401 (2016).

\bibitem{genuine13} Q. Y. He and M. D. Reid, Genuine multipartite
Einstein-Podolsky-Rosen steering, Phys. Rev. Lett. \textbf{111}, 250403
(2013).

\bibitem{Adesso15} I. Kogias, A. R. Lee, S. Ragy, and G. Adesso,
Quantification of Gaussian quantum steering, Phys. Rev. Lett. \textbf{114}%
, 060403 (2015).


\bibitem{ANUexp} S. Armstrong, M.Wang, R. Y. Teh, Q. H. Gong, Q. Y. He, J. Janousek, H. A. Bachor, M. D. Reid, and P. K. Lam, Multipartite Einstein-Podolsky-Rosen steering and genuine tripartite entanglement with optical networks, Nat. Phys. \textbf{11}, 167--172 (2015).

\bibitem{GiannisQSS}I. Kogias, Y. Xiang, Q. Y. He, and G. Adesso, Unconditional security of entanglement-based continuous-variable quantum secret sharing, Phys. Rev. A \textbf{95}, 012315 (2017).



%\bibitem{Reid89} M. D. Reid, Demonstration of the Einstein-Podolsky-Rosen
%paradox using nondegenerate parametric amplification, Phys. Rev. A \textbf{40}, 913--923 (1989).


\bibitem{prlSu} X. W. Deng, Y. Xiang, C. Tian, G. Adesso, Q. Y. He, Q. H. Gong, X. L. Su, C. D. Xie, and K. C. Peng, Demonstration of monogamy relations for Einstein-Podolsky-Rosen steering
in Gaussian cluster states, Phys. Rev. Lett. \textbf{118}, 230501 (2017).

\bibitem{Pinel}O. Pinel, P. Jian, R. M. de Ara\'{u}jo, J. Feng, B. Chalopin, C. Fabre, and N. Treps, Generation and characterization of multimode quantum frequency combs, Phys. Rev. Lett. \textbf{108}, 083601 (2012).

\bibitem{Pfister}M. Pysher, Y. Miwa, R. Shahrokhshahi, R. Bloomer, and O. Pfister, Parallel generation of quadripartite cluster entanglement in the optical frequency comb, Phys. Rev. Lett. \textbf{107}, 030505  (2011).

\bibitem{Roslund}J. Roslund, R. M. de Ara\'{u}jo, S. Jiang, C. Fabre, and N. Treps, Wavelength-multiplexed quantum networks with ultrafast frequency combs,  Nat. Photonics \textbf{8}, 109 (2014).

\bibitem{multient2015}S. Gerke, J. Sperling, W. Vogel, Y. Cai, J. Roslund, N. Treps, and C. Fabre, Full Multipartite Entanglement of Frequency-Comb Gaussian States, Phys. Rev. Lett. \textbf{114}, 050501 (2015).

\bibitem{vogel2016} S. Gerke, J. Sperling, W. Vogel, Y. Cai, J. Roslund, N. Treps, and C. Fabre, 
Multipartite Entanglement of a Two-Separable State, Phys. Rev. Lett. \textbf{117}, 110502 (2016).

\bibitem{Cainc}Y. Cai, J. Roslund, G. Ferrini1, F. Arzani, X. Xu, C. Fabre, and N. Treps, Multimode entanglement in reconfigurable graph
states using optical frequency combs,  Nat. Commun. \textbf{8}, 15645 (2017).


\bibitem{supp} See Appendix for details of the squeezing level and CMs of prepared states,
comparison with single mode squeezed state, and generation of one-way steering via asymmetric control. The Appendix contains additional references \cite{OneWayPryde,OneWayGuo,OneWayNatPhot}.

\bibitem{OneWayPryde} S. Wollmann, N. Walk, A. J. Bennet, H. M. Wiseman, and
G. J. Pryde, Observation of genuine one-way Einstein-Podolsky-Rosen
steering, Phys. Rev. Lett. \textbf{116}, 160403 (2016).

\bibitem{OneWayGuo} K. Sun, X. J. Ye, J. S. Xu, X. Y. Xu, J. S. Tang, Y. C. Wu, J. L. Chen, C. F.
Li, and G. C. Guo, Experimental quantification of asymmetric
Einstein-Podolsky-Rosen steering, Phys. Rev. Lett. \textbf{116}, 160404
(2016).

\bibitem{OneWayNatPhot} V. H\"{a}ndchen, T. Eberle, S. Steinlechner, A. Samblowski, T. Franz, R. F. Werner, and R. Schnabel, Observation of one-way Einstein-Podolsky-Rosen steering, Nat. Photonics \textbf{6}, 596 (2012).

\bibitem{Yunon}Y. Xiang, B. Xu, L. Mi\v{s}ta, Jr., T. Tufarelli, Q. Y. He, and G. Adesso, Investigating Einstein-Podolsky-Rosen steering of continuous-variable bipartite states by
non-Gaussian pseudospin measurements, Phys. Rev. A \textbf{96}, 042326 (2017).

\bibitem{Yumonogamy} Y. Xiang, I. Kogias, G. Adesso, and Q. Y. He, Multipartite Gaussian steering: Monogamy constraints and quantum cryptography applications, Phys. Rev. A \textbf{95}, 010101(R) (2017).

\bibitem{Reidmonogamy}M. D. Reid, Monogamy inequalities for the Einstein-Podolsky-Rosen paradox and quantum steering, Phys. Rev. A \textbf{88}, 062108 (2013).

\bibitem{Kimmonogamy}S-W. Ji, M. S. Kim and H. Nha, Quantum steering of multimode Gaussian states by Gaussian measurements: Monogamy relations and the Peres conjecture, J. Phys. A: Math. Theor. \textbf{48}, 135301 (2015).

\bibitem{GSmonogamy} G. Adesso and R. Simon, Strong subadditivity for log-determinant of covariance matrices and its applications, J. Phys. A: Math. Theor.  \textbf{49}, 34LT02 (2016).



\bibitem{Adesso16} L. Lami, C. Hirche, G. Adesso, and A. Winter, Schur Complement Inequalities for Covariance Materices and Monogamy of Quantum Correlations, Phys. Rev. Lett. \textbf{117}, 220502
(2016).

\bibitem{G4qubitmonogamy} B. Regula, A. Osterloh, and G. Adesso, Strong monogamy inequalities for four qubits, Phys. Rev. A \textbf{93}, 052338 (2016).

\bibitem{Shumingmonogamy} S. M. Cheng, A. Milne, M. J. W. Hall, and H. M. Wiseman, Volume monogamy of quantum steering ellipsoids for multiqubit systems, Phys. Rev. A \textbf{94}, 042105 (2016).

\bibitem{Shumingmonogamyexp} C. Zhang, S. Cheng, L. Li, Q. Liang, B. Liu, Y. Huang, C. Li, G. Guo, M. J. W. Hall, H. M. Wiseman, and G. J. Pryde, Experimental Validation of Quantum Steering Ellipsoids and Tests of Volume Monogamy Relations, Phys. Rev. Lett. \textbf{122}, 070402 (2019).


\bibitem{100mode}G. Patera, N. Treps, C. Fabre, and G. J. de Valcarcel, Quantum theory of synchronously pumped type I optical parametric oscillators: characterization of the squeezed supermodes, Eur. Phys. J. D \textbf{56}, 123 (2010).

\bibitem{separatesepctal}G. Ferrini, J.-P. Gazeau, T. Coudreau, C. Fabre, and N. Treps, Compact Gaussian quantum computation by multi-pixel homodyne detection, New J. Phys. \textbf{15}, 093015 (2013).

\end{thebibliography}
\end{document}